# Atomic-scale study on core-shell Cu precipitation in steels: atom probe tomography and ab initio calculations

Xiao Shen[1,2,#], YiXu Wang[3,#], Zigan Xu[2], Bowen Zou[1], Enzo Liotti[4], Richard Dronskowski[3], Wenwen Song[1,2,*]

[1] Institute of Materials Engineering, University of Kassel, Mönchebergstr. 3, 34125 Kassel, Germany

[2] Steel Institute, RWTH Aachen University, Intzestr. 1, 52072 Aachen, Germany

[3] Institute of Inorganic Chemistry, RWTH Aachen University, Landoltweg 1, 52056 Aachen, Germany

[4] Department of Materials, University of Oxford, Parks Road, OX1 3PH Oxford, UK

[#] These authors contributed equally to this work.

[*] Corresponding author

E-mail: song@uni-kassel.de

**Keywords:** density functional theory, atom probe tomography, nano-sized Cu precipitate, core-shell structured precipitate

**Abstract:** The present work investigates the atomic interactions among Cu, Al, and Ni elements in bcc-iron matrix, focusing on the formation mechanism of nano-sized core-shell Cu precipitates. Using a combination of atom probe tomography (APT), density functional theory (DFT) calculations, and molecular dynamics (MD) simulations, the study provides insights into the atomic-scale migration tendencies of these elements in the supersaturated solid solution surrounding Cu precipitate in the martensite phase of a medium-Mn steel. The results show that Ni and Al atoms were not expelled by Cu atoms but were instead attracted to the bcc iron matrix, forming a stable co-segregation in the outer shell. This phase effectively surrounded the nano-sized Cu precipitate and prevented its rapid growth, contributing to improved mechanical properties. The findings offer a theoretical method for developing Cu-contaminated circular steels by utilizing DFT calculations to unravel bonding preferences and assess the potential for forming a stable precipitation phase around nano-sized Cu precipitates.

## 1. Introduction

Steel production and processing significantly contribute to greenhouse gas emissions, making the development of high-performance circular steels crucial for sustainable targets. The advancement of circular steels requires a controlled balance between optimizing alloying elements to improve mechanical properties and mitigating the detrimental effects of impurities, often termed tramp elements. Cu exhibits a two-fold role in steel development. On the one hand, Cu is frequently considered as a problematic tramp element in circular steels, which is difficult to remove and can segregate during semi-solid processing, leading to issues like grain boundary embrittlement and hot shortness even at low concentrations [1]. On the other hand, with proper alloy design, Cu may contribute significantly to steel strengthening through mechanisms of solid solution strengthening and precipitation hardening [2,3,4,5,6].

The effects of Cu in steels are neither exclusively negative nor uncontrollable. Our previous study has shown an effective strategy for managing Cu alloying to enhance damage tolerance and tailoring of the mechanical properties in different grades of steels [7,8,9,10,11,12]. Shen *et al.* [8] reported the influence of C concentration on microstructural evolution and Cu precipitation in aged ultralow and medium C ferritic steels, *i.e. X0.5CuNi2-2 and X21CuNi2-2*. The elevated C concentration led to a larger nucleation density, heterogeneous nuclei distribution upon the onset of aging and further rapid growth in a large precipitate size. The size, distribution and coherency largely influenced the hardness and cyclic hardening potential in the investigated steels. Görzen *et al.* [7] further evaluated the defect tolerance of the same group of ferritic steels after aging treatment. An increase in aging time was accompanied by an increase in defect tolerance for X0.5CuNi2-2 steel, which correlates with a higher cyclic hardening potential induced by the formation of Cu precipitates. Cu alloying in medium-Mn steels (MMnS) was also reported positive in tailoring austenite stability and consequently the mechanical properties. The co-additions of 1.5 wt.% Cu and Ni into Fe-0.05C-7Mn-1.5Al-1.5Si-0.5Mo (wt.%) steel contributed to enhanced thermal stability of austenite, preventing the athermal martensitic transformation during quenching [11]. The Cu, Ni and Al alloying triggered the core-shell structured Cu precipitation, with Ni and Al enriched in an outer shell surrounding the nano-size Cu particle (core). The significant increase (197 MPa) in yield strength of the investigated Cu-Ni alloyed MMnS was attributed to the strong interaction between mobile dislocation and nano-sized Cu precipitate [11,12].

Although Ni+Al addition has been considered as an effective and empirical solution to mitigate the Cu-induced hot shortness in steels, the mechanisms of how these elements (Cu, Al, Ni) interact with each other and with matrix Fe, and how they promote the complex core-shell precipitation are not fully understood. Zhang and Enomoto's [13] study has shown that Ni tends to accumulate at some distance from the center of Cu nucleus, reducing the work for the formation of a Cu critical nucleus. However, the understandings of early stage of Cu atoms clustering and how Ni, Al atoms are accumulated away from Cu nucleus still remain blank. A deep understanding of nano-sized Cu precipitation mechanism and atomic interaction is significant in developing Cu-management strategies for circular steels.

In the authors' previous work [11] on the investigated Cu-alloyed medium-Mn steel (MMnS), Fe–0.05C–7Mn–1.5Cu–1.5Ni–1.5Al–1.5Si–0.5Mo (wt.%), the microstructure, morphology, and grain size were thoroughly investigated using electron backscattering diffraction technique (EBSD). Moreover, high-energy synchrotron X-ray diffraction was employed to accurately compare the austenite fraction in the non-deformed state and after necking, providing precise insights into phase transformation behavior under mechanical loading. In the present work, we focus on studying the atomic interactions among alloying elements (Cu, Al, Ni) and matrix Fe using a combined experimental approach and DFT calculations to interpret the formation mechanism of nano-sized core-shell Cu precipitate. This fundamental understanding will further stimulate alloying and processing strategies for the applications of Cu-contaminated circular steels.

## 2. Materials and methods

The Cu nano-precipitate was characterized on a Cu-alloyed medium-Mn steel. The nominal composition of the investigated steel is Fe–0.05C–7Mn–1.5Cu–1.5Ni–1.5Al–1.5Si–0.5Mo (wt.%). The investigated material, cast at the Steel Institute (IEHK) of RWTH Aachen University in a laboratory vacuum induction furnace, was homogenized at 1250°C for 8 hours, then hot forged at 900-1100°C, reducing the cross-section to 100 mm x 100 mm. The sample was heat-treated by a short annealing + tempering process (SAT), *i.e. heated to 700 °C and isothermally held for 2 minutes, water-quenched to room temperature, subsequently followed by a tempering at 500 °C for 3 hours before water quenching to room temperature.* The atom probe sample was prepared using the FEI Helios Nanolab 660 dual beam FIB-SEM system. The three-dimensional (3D) characterization of Cu nano-precipitates and local chemistry were performed by a Local Electron Atom Probe (LEAP) 4000X HR system (CAMECA Instrument Inc.). Laser-pulse mode (wavelength of 355 nm, frequency of 200 kHz, laser energy of 30 pJ, detection rate of 0.5%) was selected for data collection. The base temperature in the analysis chamber was kept at 60 K during the measurement. The collected data were reconstructed and analyzed using the software AP suite 6.1 (CAMECA Instruments Inc.).

The DFT calculations were performed using the Vienna Ab initio Simulation Package [14] (VASP, version: 6.1.1). The projector augmented wave (PAW) pseudopotentials with Perdew–Burke–Ernzerhof (PBE) generalized gradient approximation (GGA) treating exchange-correlation functionals [15]. To model the solid solution, a special quasi-random structure (SQS) [16] was generated by icet [17,18] in a 2 × 2 × 2 supercell of the primitive cell of body center cubic (bcc) structure. The DFT calculations focus on the region around the interface between Cu precipitate and the Fe matrix, where equimolar of those four elements (Fe, Al, Ni, Cu) was assumed. A Γ-centered Monkhorst–Pack 3 × 3 × 3 $k$-point grid was generated [19] for sampling the first Brillouin zone and the tetrahedron method with Blöchl corrections was used [20]. The energetic criterion for convergence was set at $10^{-8}$ eV. Spin-polarization was considered and the starting magnetic moments for Fe and Ni were set at 5 $\mu_B$. Molecular dynamic (MD) simulations at 800 K were performed for 2 ps following the NVT ensemble, with a step of 1 fs. The chemical bonding information was extracted by using the Local-Orbital Basis Suite Towards Electron-Structure Reconstruction [21,22] (LOBSTER, version: 5.0.0 [23]). Crystal Orbital Hamilton Population (COHP) [24] analyses were performed on every bond of the bcc structure up to the fourth nearest neighbor.

## 3. Results and discussions

### 3.1. Atom probe tomography (APT) characterization

In the current study, a detailed APT analysis was performed to characterize the core-shell structured nano-sized Cu precipitate in martensite phase in the investigated Fe–0.05C–7Mn–1.5Cu–1.5Ni–1.5Al–1.5Si–0.5Mo (wt.%) steel at near-atomic scale. **Figure 1** a) shows the reconstructed APT tip (martensitic matrix) with a large number density of nano-sized Cu precipitates. The average radius of nano-sized Cu precipitates (rendered by 6 at.% Cu iso-concentration surface) was calculated as 2.95 nm with a high number density of 9.5 x $10^{22}$ m$^{-3}$. The weight concentrations of C, Mn, Cu, Ni, Al, Si, Mo in martensitic matrix (excluding 6 at.% Cu precipitate) were determined as 0.005 wt.%, 3.7046 wt.%, 0.5394 wt.%, 1.2874 wt.%, 1.9070 wt.%, 2.0674 wt.% and 0.5535 wt.% respectively. A representative chemical composition of nano-sized Cu precipitate is displayed in Figure 1 b). The difference in evaporation fields of different elements can lead to local magnification effects, which have been reported to introduce uncertainties in the analysis of precipitates and clusters [25,26]. In this context, for APT analyses of nano-sized Cu precipitates in the present work, the z-direction (analysis direction) of the cylindrical region of interest (ROI) was consistently aligned with the z-direction of the APT dataset to minimize bias. A clear core-shell structured nano-sized Cu precipitate shows the chemical complexity in the core and shell regions, respectively. Figure 1 c) shows a cross-sectional view obtained by cropping different iso-concentration surfaces, illustrating that the nano-sized Cu precipitate is fully encapsulated by a Ni- and Al-enriched shell. In the core region, the highest Cu concentration of ~45 at.% is observed with a significant depletion of Fe content. The Cu concentration doesn't approach 100 at.%, which indicates the early stage of Cu precipitate growth. Considering the relatively long tempering duration, the growth of Cu precipitate has been largely impeded in comparison with the authors' previous study on Cu precipitate evolution as the function of aging duration in ferritic steels [8]. In the shell region, co-enrichment of Ni and Al is observed, and the atomic ratio of Ni/Al is maintained at approximately 1:1. This finding indicates the co-segregation of Ni and Al forms at Cu precipitate-matrix interface, which has been reported to retard the growth of nano-sized Cu precipitates [27,28]. To maintain the Cu precipitates at the nanoscale, the stabilization of the shell region through Ni and Al enrichment, followed by the formation of the NiAl phase, has proven to be an effective strategy [27]. However, the mechanism by which Ni and Al accumulate around the Cu precipitates remains unclear.

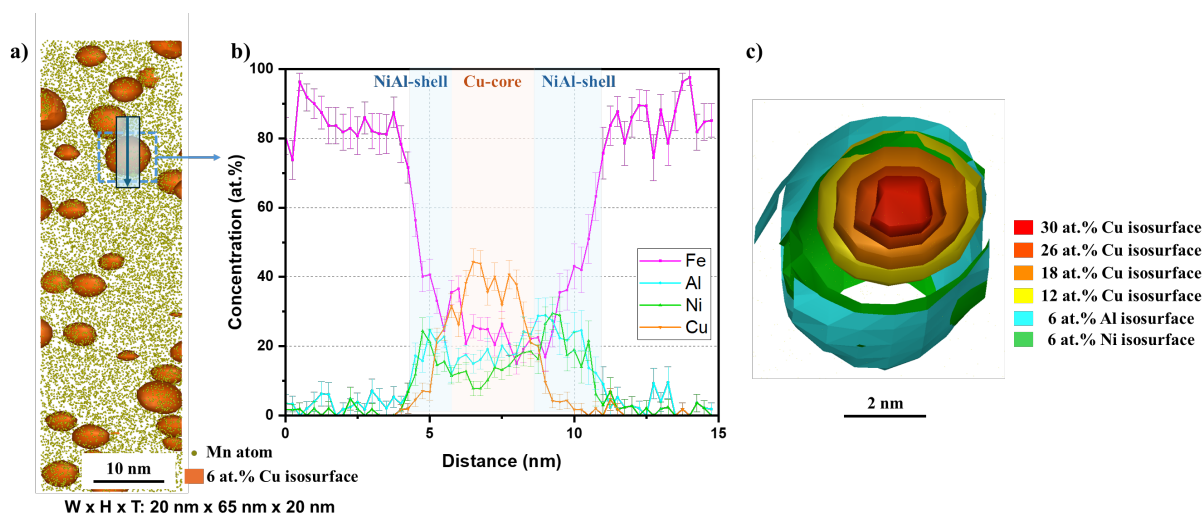

**Figure 1** APT analysis of core-shell structured Cu precipitate for SAT-processed MMnS, Fe–0.05C–7Mn–1.5Cu–1.5Ni–1.5Al–1.5Si–0.5Mo (wt.%): a) reconstructed APT cut (20 nm x 65 nm x 20 nm) showing large density of nano-sized Cu precipitate; b) Chemical profiles of Fe, Al, Ni, Cu elements in the selected Cu precipitate (as marked by the blue dash line in a)), showing the core-shell structure covered by Ni, Al enrichment; c) Cross-sectional view of the nano-sized Cu precipitate fully covered by co-enrichment of Ni and Al.

### 3.2. Crystal Orbital Hamilton Population (COHP) analysis of the entire system

The cursory look on chemical bonds for the entire system can be taken by averaging the Crystal Orbital Hamilton Population (COHP) for all the same kind of chemical bonds with the same bond length (see **Figure 2**). For comparison, interactions between the same elements were grouped into the same subplot. Obviously, judging from projected COHP (pCOHP), the interactions for the third- and fourth-nearest neighbors are significantly smaller than those for the first- and second-nearest neighbors. This is also mirrored by the integrated pCOHP (IpCOHP) values, which were deduced by integrating the COHP curve to reflect the bond strength. The IpCOHP values for the third- and fourth-nearest neighbor of a type of bond are usually close to only about 10% of the values for the first- and second-nearest neighbors. Hence, the trivial influences from the third- and fourth-nearest neighbors were expected. The strongest bond in the system is Al–Al (–1.429 eV/bond) whereas the weakest one is Cu–Cu (–0.164 eV/bond). All IpCOHP values have been shown in **Figure 3** a).

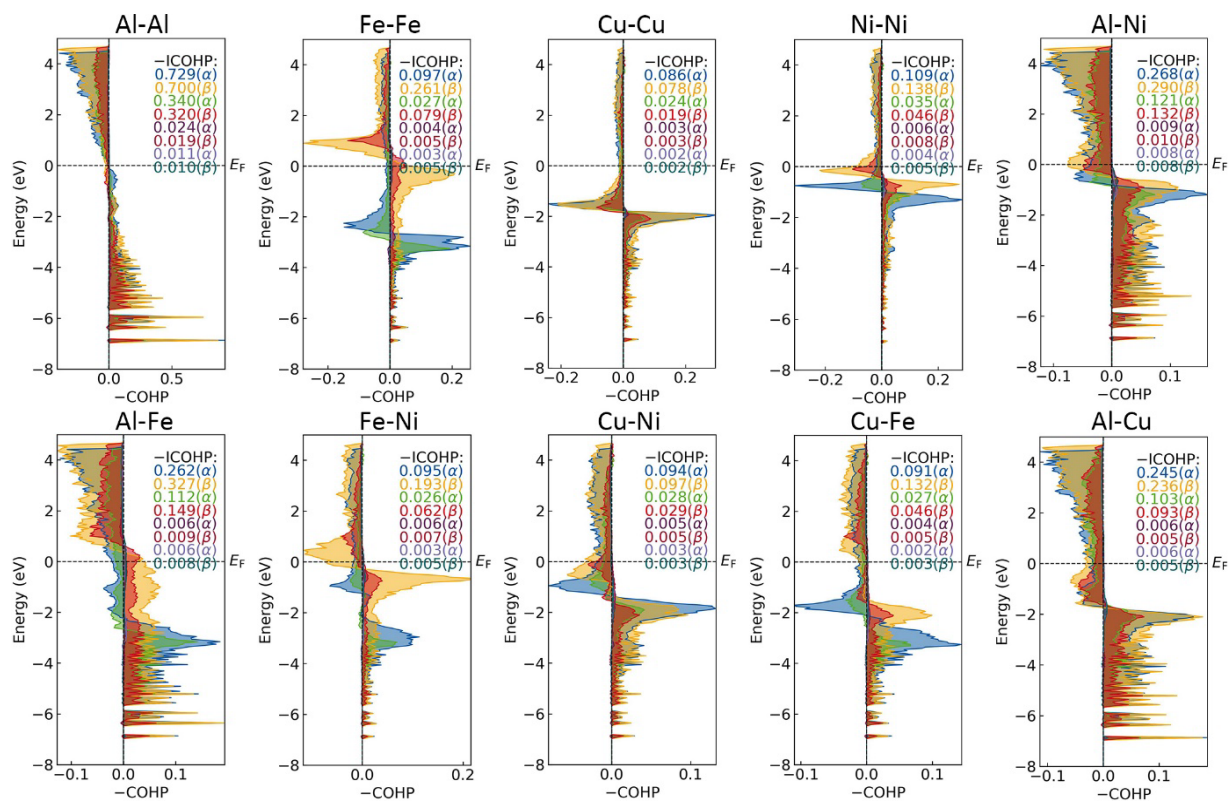

**Figure 2** COHP analysis of chemical bonds in the SQS model. From the upper right corner down, each pair of α and β denotes the majority/minority spin for first-, second-, third-, and fourth-nearest neighbor interactions, respectively. The integrated values were also provided to reflect the bond strengths. Those values are colored in the same color as the corresponding COHP curve. Note that negative values are plotted such as to have all bonding contributions go to the right, a simple graphical convention.

It is worth noting that a larger bond strength does not necessarily indicate a more favorable bond formation as the evolution of a system is towards elimination of antibonding levels near the Fermi level [29]. Comparing all homologous bonds, the position of the Fermi level resides almost at the non-bonding/bonding region, except for the Ni–Ni bond, for which the Fermi level is located at the antibonding peaks for the spin-minority ($\beta$) channel (because of spin-polarization) of the first- and second-nearest neighbor interactions. Therefore, forming Ni–Ni bond in such a system is not favorable whereas the formation of Al–Al, Cu–Cu, and Fe–Fe is relatively favored. Given fewer of those bonds in the system, the influence of those chemical bonds may be limited but useful to interpret the precipitation of Ni–Al. In the same way, it can also be anticipated that the formation of Al–Ni, Al–Cu, and Cu–Ni is not favored. On the contrary, the formation of Cu–Fe, Fe–Ni and Al–Fe bonds is relatively favorable. Hence, we would expect that there will still be significant interactions between the precipitates and the matrix. The Ni–Al core surrounding the Cu precipitates would be the consequences of those interactions. We have dived deeper into the local chemical environment and found that the findings still hold, please refer to the supporting information.

### 3.3. Formation Mechanism of NiAl Shell Surrounding the Cu Nuclei

Admittedly, COHP analysis is hinged on a static DFT calculation. It is hard to accurately predict atomic movement solely based on COHP results, even though we already got hints on the preferences of bond formation. Herein, we would interpret such "preferences" as the tendency to form a chemical bond at the corresponding equilibrium distances. Activated by the thermal fluctuation at finite temperatures, the atoms would become more mobile and the movement of atoms would be dominated by the well-known Hellmann-Feynman force (depending on the distribution of electrons within the systems) on the level of DFT. Therefore, we hypothesized that there would be a different arrangement of atoms due to populated anti-bonding states in the system [30]. MD results are in good agreement with the conclusions from COHP studies. The radial distribution function (RDF) has been displayed in Figure 3 b). Compared with other bonds, which are essentially amorphous and only present one well-defined peak corresponding to the nearest neighbor interactions, it can be clearly observed that Cu atoms are clustered as the position of its first peak (2.5 Å) corresponds to the nearest neighbor distance of fcc Cu-Cu bond (2.53 Å) and the tiny peak in the vicinity of 3.5 angstrom is very close to the second nearest neighbor of fcc Cu-Cu bond (3.58 Å). These findings suggest that clusters of Cu atoms started to form and NiAl started to co-enrich simultaneously.

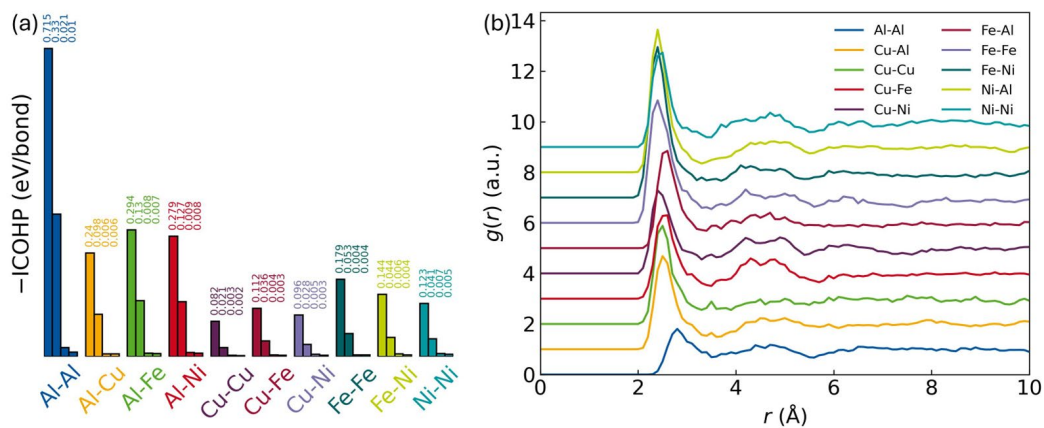

**Figure 3** (a) Histogram of IpCOHPs for various chemical bonds in the SQS model. Within the same kind of bond, the bar at the very left position corresponds to the first-nearest neighbor and the one at the very right position corresponds to the fourth-nearest neighbor. (b) Radius distribution function of pairwise interactions in the system after 2 ps of molecular dynamic simulation.

Based on the experimental observation, DFT calculations and MD simulation in the shell region, we can elucidate the formation mechanism of nano-sized core-shell Cu precipitates from the perspective of atomic bonding tendencies. From the COHP analysis of the entire system, Fe–Al and Fe–Ni bonds are more favored to form, which is also validated in the COHP analysis of the local chemical environment. Therefore, in a system with randomly distributed atoms (assumed in the outer layer of Cu cluster), Fe–Al and Fe–Ni bonds preferentially form, attracting Ni and Al elements towards matrix-Fe atoms. Ni and Al atoms tend to enrich outward in the shell region, moving closer to the Fe matrix. Although Fe–Cu bonds may also form, their bonding tendency is weaker than Fe–Al and Fe–Ni bonds. Only Ni and Al atoms accumulate near precipitate-matrix interface, potentially leading to the formation of B2-typed NiAl phase. In the region closely adjacent to the Ni and Al-enriched zone, due to the migration of Ni and Al atoms, COHP analysis indicates that the remaining Cu atoms bond with Cu atoms in the precipitate core, forming Cu–Cu bonds. As suggested by the COHP analysis, this atomic movement leads to the formation of the core-shell structured Cu precipitate with NiAl. It is also worth discussing that, according to the COHP analysis, if in the absence of Al in the alloy, Ni atoms tend to segregate at Cu precipitate-matrix interface and form a shell structure. However, due to the instability of Ni–Ni bonds, the Fe–Ni interactions cause the Ni shell to decompose into the Fe matrix. This inference is consistent with our previous work of ferritic steels (X0.5CuNi2-2 and X21CuNi2-2) [8]. Due to the very low Al content (0.04 wt.%), no Ni enrichment at the interface between the Cu precipitates and the Fe matrix was observed during the growth stage of the Cu precipitates [8].

## 4. Conclusions

The present work employed DFT calculations and MD simulations to investigate the atomic-scale migration tendencies of Ni, Al, and Cu elements in the supersaturated solid solution surrounding Cu precipitate. The results confirmed the core-shell structure of nano-sized Cu precipitates observed in APT experiments. The results reveal a notable deviation from previous understanding: Ni and Al atoms are not expelled by Cu atoms to form the shell region. Instead, the supersaturated Ni and Al atoms in the outer layer of the Cu precipitate are attracted by the Fe atoms in the bcc-matrix and migrate outwards, forming co-segregation. This work provides a theoretical approach for designing Cu-contaminated circular steels. By using DFT to analyze the bonding preferences of different elements with the matrix, it is possible to predict how to form a stable phase at the Cu precipitate shell. The stable phase as shell region covers the nano-sized Cu precipitate and may act as an obstacle to prevent the fast growth of nano-sized Cu precipitate, which may consequently enhance the mechanical performance of steels.

**Acknowledgements**

The authors, X.S., W.S. and B.Z. gratefully acknowledge the Deutsche Forschungsgemeinschaft (DFG) for funding the research work (grant number BL 402/49-1 and SO 1431/7-1). X.S. and Y.W. contributed equally to this work.

Received: ((will be filled in by the editorial staff))
Revised: ((will be filled in by the editorial staff))
Published online: ((will be filled in by the editorial staff))

ToC figure

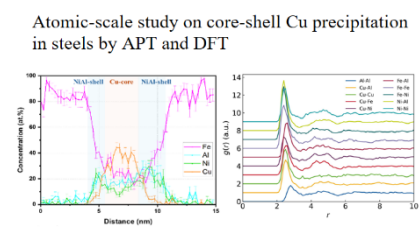